\documentclass[aps,pra,amsmath,amssymb,preprint,letterpaper,superscriptaddress,11pt]{revtex4-1}

\usepackage{hyperref}
\usepackage{graphicx}

\newcommand{\trans}{\mathsf{T}}

\begin{document}

\title{Multi-component symmetry-projected approach for molecular
  ground state correlations}

\author{Carlos A. Jim\'enez-Hoyos}
\affiliation{Department of Chemistry, Rice University, Houston, TX
  77005}

\author{R. Rodr\'iguez-Guzm\'an}
\affiliation{Department of Chemistry, Rice University, Houston, TX
  77005}
\affiliation{Department of Physics and Astronomy, Rice University,
  Houston, TX 77005}

\author{Gustavo E. Scuseria}
\affiliation{Department of Chemistry, Rice University, Houston, TX
  77005}
\affiliation{Department of Physics and Astronomy, Rice University,
  Houston, TX 77005}

\date{\today}

\begin{abstract}
The symmetry-projected Hartree--Fock ansatz for the electronic
structure problem can efficiently account for static correlation in
molecules, yet it is often unable to describe dynamic correlation in a
balanced manner. Here, we consider a multi-component,
systematically-improvable approach, that accounts for all ground state
correlations. Our approach is based on linear combinations of
symmetry-projected configurations built out of a set of
non-orthogonal, variationally optimized determinants.  The resulting
wavefunction preserves the symmetries of the original Hamiltonian even
though it is written as a superposition of deformed (broken-symmetry)
determinants. We show how short expansions of this kind can provide a
very accurate description of the electronic structure of simple
chemical systems such as the nitrogen and the water molecules, along
the entire dissociation profile. In addition, we apply this
multi-component symmetry-projected approach to provide an accurate
interconversion profile among the peroxo and bis($\mu$-oxo) forms of
[Cu$_2$O$_2$]$^{2+}$, comparable to other state-of-the-art quantum
chemical methods.
\end{abstract}

\maketitle

\section{Introduction}
\label{sec:introduction}

In recent work
\cite{jimenez-hoyos2012,samanta2012,garza2013,rivero2013a,rivero2013b},
we have explored the merits of the symmetry-projected Hartree--Fock
(HF) ansatz for describing the electronic structure of molecular
systems. A symmetry-projected ansatz can account for most of the
static correlations present in molecular systems while also capturing
a fraction of the dynamic correlation. The resulting wavefunction has
a highly non-trivial multi-determinantal character, with well defined
quantum numbers, and yet it is described by a single set of occupied
orbitals. In this way, the connection to the single-particle picture
is not completely abandoned. The projected state can be expressed as
the resonance among the different broken-symmetry, defect-possessing
Slater determinants, in such a way that a state with well-defined
symmetries is recovered.

A successful many-body approach to quantum chemistry must be able to
predict reaction energies and reaction barriers with (near) chemical
accuracy. That is, the method must be able to predict energy
differences between reactants, products, and transition states to
within a few kcal/mol, which is a small fraction of the total
electronic energies. In order to accomplish this task for general
chemical systems, the method must provide a balanced description of
static and dynamic correlations for the different chemical species
participating in a given process.  A symmetry-projected HF approach
will generally fail these tests: the method is not size consistent and
the amount of correlations captured is, to a given extent, system and
symmetry dependent.

In this work, we explore a systematic way to approach the exact
many-body wavefunction by taking linear combinations of
symmetry-projected configurations. Ideally, the multi-component
approaches here considered should account for most of the correlations
(both static and dynamic) in chemical systems with just a few
symmetry-projected configurations. If the number of such
configurations depends weakly on the size of the system, the approach
remains mean-field in computational cost.

Our multi-component approach follows the few-determinant (FED)
treatment described by Schmid \cite{schmid1989,schmid2004} in the
nuclear physics community as well as the resonating HF (Res HF)
approach originally proposed by Fukutome \cite{fukutome1988}. The two
constitute extreme strategies of a more general method where linear
combinations of symmetry-projected configurations are used, regardless
of the approach used to optimize them. We note that a linear
combination of restricted HF determinants was used by Koch and
Dalgaard \cite{koch1993} to reach near full configuration-interaction
(FCI) accuracy in the electronic energies of Be, BH, and
H$_2$O. Similarly, a Res HF approach was used by Tomita, Ten-no, and
Tanimura \cite{tomita1996} in half-projected calculations on carbon
monoxide.  Both FED and Res HF approaches based on symmetry-projected
configurations have been successfully applied to describe
strongly-correlated systems in condensed matter physics like the one
and two-dimensional Hubbard model
\cite{rodriguez-guzman2013,tomita2009,tomita2004,ikawa1993}.  In
general terms, the methods described in this paper fall within the
category of non-orthogonal configuration interaction approaches of
which there are several examples in quantum chemistry
\cite{broer1988,hollauer1993}.

This work is organized as follows. In Sec. \ref{sec:formalism} we
provide details of our formalism. In particular, after a brief review
of the symmetry-projected HF approach, we describe the general
multi-component approach and then briefly consider the FED and Res HF
strategies to optimize the resulting ansatz. In section
\ref{sec:details} we describe some features of our computational
implementation. In Sec. \ref{sec:discussion}, we apply the
multi-component formalism to describe the correlation in the
dissociation profile of N$_2$ and H$_2$O. We have also considered the
challenging [Cu$_2$O$_2$]$^{2+}$ species with our new
approach. Lastly, Sec. \ref{sec:conclusions} is devoted to concluding
remarks and work perspectives.

\section{Formalism}
\label{sec:formalism}

In this section, we describe in detail the formalism we use. We
consider the symmetry-projected HF ansatz for the ground state of a
molecular system in Sec. \ref{ssec:sphf}. We describe a CI expansion
based on symmetry-projected configurations in Sec. \ref{ssec:ciexp},
which we actually do not use but let us nicely put in perspective our
multi-component approach. Lastly, in Sec. \ref{ssec:mc} we describe
the multi-component approach, focusing in the FED and Res HF
strategies used in its optimization.

\subsection{Symmetry-projected Hartree--Fock}
\label{ssec:sphf}

We start this section by clarifying that we understand a
symmetry-projected ansatz as a wavefunction where good quantum numbers
are restored from a broken-symmetry state even if ``true'' projection
operators (in the strict mathematical sense) are not used
\cite{ring_schuck}. The symmetry-projected HF ansatz takes the form
\cite{rodriguez-guzman2012}
\begin{equation}
  |\Psi_{j,m} \rangle = \sum_k f_k \, \hat{P}^j_{mk} |\Phi \rangle,
  \label{eq:sphf_ansatz}
\end{equation}
where $\hat{P}^j_{mk}$ is a ``projection-like'' operator (written for
general non-Abelian groups) and $\{ f \}$ is an expansion of linear
variational coefficients. The subscripts $j,m$ in $|\Psi \rangle$
label the irreducible representation and the row of the irrep that are
recovered, respectively. The linear combination among different
components of the irreducible configuration is used in order to remove
an unphysical dependence of the resulting state on the orientation of
the broken-symmetry determinant \cite{schmid1984,ring_schuck}.

The symmetry-projected HF ansatz has a long history in quantum
chemistry. Originally proposed by L\"owdin in 1955, it was usually
associated with spin projection out of an unrestricted reference
determinant \cite{lowdin1966,mayer1980}. Only recently our research
group has shown \cite{jimenez-hoyos2012}, borrowing techniques
commonly applied in the nuclear physics community, how to efficiently
carry out the fully-variational optimization of symmetry-projected HF
configurations. Our strategy is based on using all symmetries of the
molecular Hamiltonian, including those that are not spontaneously
broken, and projecting them in a fully self-consistent variational
approach.

The projection operators we use take the generic form
\begin{equation}
  \hat{P}^j_{mk} = \frac{1}{V} \int_V d\vartheta \, w^j_{mk}
    (\vartheta) \, \hat{R} (\vartheta).
  \label{eq:gen_proj}
\end{equation}
Here, $\vartheta$ labels the elements of the symmetry group; for
discrete groups (such as most point groups), the integration should be
understood as a summation. In addition, $V$ is the volume of
integration, $w^j_{mk} (\vartheta)$ is an integration weight
(character) associated with the symmetries of the state to be
recovered, and $\hat{R} (\vartheta)$ is a rotation operator. We point
the interested reader to Refs. \onlinecite{ring_schuck,schmid2004} for
more details of the form of the projection operators. Given the form
(Eq. \ref{eq:gen_proj}) of the projection operator, the
symmetry-projected HF wavefunction can be expressed as a superposition
of states of the form $\hat{R} (\vartheta) |\Phi \rangle$, that is,
all degenerate states (the Goldstone manifold) generated by the set of
operators commuting with the Hamiltonian \cite{blaizot_ripka}. The
coefficients in the linear expansion $w^j_{mk} (\vartheta)$ are fully
determined by the irrep to be recovered.

The energy of the ansatz of Eq. \ref{eq:sphf_ansatz} is given by
\begin{align}
  E_j [\Phi]
  &= \, \frac{\sum_{kk'} f_k^\ast \, f_{k'} \, \langle \Phi|
    \hat{P}_{mk}^{j\dagger} \, \hat{H} \, \hat{P}^j_{mk'} |\Phi
    \rangle}{\sum_{kk'} f_k^\ast \, f_{k'} \, \langle \Phi|
    \hat{P}_{mk}^{j\dagger} \, \hat{P}^j_{mk'} |\Phi \rangle}
  \nonumber \\[4pt]
  &= \, \frac{\sum_{kk'} f_k^\ast \, f_{k'} \, \langle \Phi| \hat{H}
    \, \hat{P}^j_{kk'} |\Phi \rangle}{\sum_{kk'} f_k^\ast \, f_{k'} \,
    \langle \Phi| \hat{P}^j_{kk'} |\Phi \rangle}
  = \frac{\sum_{kk'} f_k^\ast \, f_{k'} \,
    \mathcal{H}_{kk'}}{\sum_{kk'} f_k^\ast \, f_{k'} \,
    \mathcal{N}_{kk'}}.
  \label{eq:sphf_energy}
\end{align}
The matrix elements appearing in Eq. \ref{eq:sphf_energy} can be
efficiently evaluated using the formulas provided in Appendix
\ref{sec:matrix_elements}. The corresponding derivation of the matrix
elements can be found in, {\em e.g.}, Ref. \onlinecite{jimenezthesis}.
For a detailed discussion of how the ansatz of
Eq. \ref{eq:sphf_ansatz} is optimized with respect to the set of
linear variational coefficients $\{ f \}$ and with respect to the
underlying broken symmetry determinant $|\Phi \rangle$ we refer the
reader to Refs. \onlinecite{jimenez-hoyos2013,jimenezthesis}. We
stress that the optimization method that we follow is different from
the one used in Ref. \onlinecite{jimenez-hoyos2012}, where a
parametrization based on the density matrix of the deformed
determinant was used. We note that a stationary point is found when
the following equations are all satisfied
\begin{align}
  \sum_{kk'} f_k^\ast \, f_{k'} \, \mathcal{N}_{kk'} &= \,
  \delta_{kk'} \\[4pt]
  \sum_{k'} \left( \mathcal{H}_{kk'} - E \, \mathcal{N}_{kk'} \right)
  f_{k'} &= \, 0 \qquad \forall \qquad k, \\
  \frac{\displaystyle \sum_{kk'} f_k^\ast \, f_{k'} \, \langle
    \Phi^a_i | \left( \hat{H} - E \right) \, P^j_{kk'} | \Phi
    \rangle}{\sum_{kk'} f_k^\ast \, f_{k'} \, \mathcal{N}_{kk'}} &= \,
  0 \qquad \forall \qquad i,a. \label{eq:brillouin}
\end{align}
Here, $E$ is the energy of Eq. \ref{eq:sphf_energy}; $i$ ($a$) is used
for a hole (particle)-coefficient in the third equality. The second
equation determines a generalized eigenvalue problem for the
coefficients $\{ f \}$ subject to the orthonormality constraint
expressed by the first equation. The last equation constitutes the
generalized Brillouin condition decoupling the ground-state solution
from excited particle-hole configurations.

\subsection{Configuration Interaction based on symmetry-projected determinants}
\label{ssec:ciexp}

Conceptually, the simplest approach to account for missing
correlations in the symmetry-projected HF ansatz is to consider a
configuration interaction approach. A full configuration interaction
ansatz can be written as
\begin{equation}
  |\Psi_{j,m} \rangle = \sum_k \hat{P}^j_{mk}
  \left( f_{0;k} |\Phi \rangle
  + \sum_{ia,k} f_{ia;k} |\Phi^a_i \rangle
  + \sum_{ijab,k} f_{ia,jb;k} |\Phi^{ab}_{ij} \rangle
  + \cdots \right)
  \label{eq_mrgs:fullci}
\end{equation}
In the above expression, we have used indices $i$ and $j$ to denote
occupied (hole) states in the broken-symmetry determinant $|\Phi
\rangle$, whereas indices $a$ and $b$ are used for unoccupied
(particle) states. The state $|\Phi_i^a \rangle$ constitutes a
singly-excited determinant out of the reference Fermi vacuum $|\Phi
\rangle$. The linear variational coefficients $f$ can be determined by
the solution to a generalized eigenvalue problem among all
configurations. We note that the above representation of the Hilbert
subspace with the appropriate symmetry is overcomplete.

Including only singly excited configurations (of the form $|\Phi_i^a
\rangle$) in the configuration-interaction expansion will in general
not lead to any improvement in the ground state energy. In particular,
the generalized Brillouin condition that a variationally optimized
symmetry-projected HF state satisfies is given by
Eq. \ref{eq:brillouin}, which makes singly-excited configurations
orthogonal to the symmetry-projected HF state through the Hamiltonian
when using the $f_{0;k}$ variational coefficients. Note that if the
dimension of the irreducible representation associated with the
restored symmetry is larger than 1, some energy improvement in the
ground state may be obtained by diagonalization in the singly-excited
space due to the variational coefficients $f_{ia;k}$. On the other
hand, diagonalization in this space can be used for a first-order
description of excited states. This correponds to the
symmetry-projected Tamm-Dancoff Approximation discussed by Schmid {\em
  et al}. \cite{schmid1989}.

If, in addition, one includes doubly-excited configurations, an energy
improvement is all but guaranteed unless the symmetry-projected HF
state was already exact. Nevertheless, the matrix is large and dense;
the evaluation of each matrix element is significantly more expensive
than in the standard HF-based approach, where the Slater--Condon rules
\cite{szabo_ostlund} can be used to simplify the evaluation.

We have not pursued the configuration interaction approach described
in this section. We strive to obtain wavefunctions as compact as
possible, in an efficient manner, in terms of a small number of
symmetry-projected configurations. This facilitates, at the same time,
the physical insight behind the wavefunction. If the ground state
wavefunction is expanded in terms of a few non-orthogonal,
symmetry-broken Slater determinants, one may relate the correlation
thus gained to the resonance among these different
configurations. This spirit is also pursued in the generalized
multistructural wavefunction of Hollauer and Nascimento
\cite{hollauer1993}, where a linear combination of expansions based on
non-orthogonal Slater determinants is used as a variational ansatz.

\subsection{Multi-component approaches}
\label{ssec:mc}

Let us suppose that we have already optimized a symmetry-projected HF
configuration. In this section, we write this as
\begin{equation}
  | \, {}^1 \Psi_{j,m} \rangle = \sum_k f^1_k \, \hat{P}^j_{mk} | \,
  {}^1\Phi \rangle,
\end{equation}
where the superscript $1$ in $|^1 \Psi_{j,m} \rangle$ is used to
indicate that a single symmetry-projected configuration is used in the
ansatz. Similarly, the superscript $1$ in the $f$ variational
coefficients and in $|\Phi \rangle$ indicate that only one determinant
is included in the ansatz.

In the general case, we can describe the ground state with $n$
symmetry-projected configurations, as in
\begin{equation}
  | \, {}^n \Psi_{j,m} \rangle = \sum_k \hat{P}^j_{mk} \sum_{l=1}^n
    f^l_k \, | \, {}^l \Phi \rangle.
  \label{eq_mrgs:ndet}
\end{equation}

Note that this defines a systematically-improvable approach. That is,
if $n$ determinants prove insufficient to provide an accurate
description of a given system, one can add one (or a few) more
configuration(s) to the expansion. It is also important to note that
the expansion of Eq. \ref{eq_mrgs:ndet} is written as a superposition
of the Goldstone manifolds associated with each of the $n$-deformed
determinants in the multi-component wavefunction. One has now to
address the issue of how to variationally optimize the
$n$-configuration ansatz. There are two extreme approaches that we
will consider:

\begin{itemize}
  \item In the FED (few-determinant) approach
    \cite{schmid1989,schmid2004,rodriguez-guzman2013}, the different
    configurations are optimized one-at-a-time. That is, the second
    symmetry-projected configuration is optimized after the first one,
    leaving the latter untouched, and so on.

  We note that there is no need for the FED expansion to be short, as
  its name would imply, although it is a desirable feature. We keep
  the acronym to remain consistent with the literature.

  \item In the resonating HF (Res HF) approach \cite{fukutome1988},
    all the configurations are optimized at the same time.
\end{itemize}

There are, of course, a number of possible variants in between. For
instance, one could optimize two configurations at a time. Each
approach has strengths and drawbacks. In particular, we would like to
stress the following:
\begin{itemize}
  \item A Res HF optimized wavefunction is stationary with respect to
    changes in any of the underlying determinants. On the other hand,
    a FED optimized wavefunction is stationary only with respect to
    particle-hole excitations of the last-added determinant.

  This feature makes the Res HF wavefunction easier to work with for
  evaluating properties that depend on derivatives of the
  wavefunction.

  \item In a Res HF optimization, $\mathcal{O} (n^2)$ overlap and
    Hamiltonian matrix elements need to be re-computed at every
    iteration. In contrast, an efficient implementation of the FED
    approach requires only $\mathcal{O} (n)$ overlap and Hamiltonian
    matrix elements to be recomputed.

  \item The convergence properties of the two approaches can be very
    different. In the Res HF approach, for instance, there is no
    guarantee that any of the configurations will resemble the
    optimized single-configuration ansatz.
\end{itemize}

The matrix elements appearing in the evaluation of the energy and
energy gradient with multi-component approaches can be efficiently
evaluated using the expressions provided in Appendix
\ref{sec:matrix_elements}. In our calculations, we carry out the
optimization with respect to the broken-symmetry determinants in the
multi-component expansion using a robust, Thouless-based
parametrization \cite{mang1975,jimenez-hoyos2012b}. We now proceed to
consider each of the two approaches in detail.

\subsubsection{The few-determinant (FED) approach}

In the few-determinant (FED) approach introduced by Schmid
\cite{schmid1989,schmid2004}, only the last-added symmetry-projected
configuration is optimized with respect to the underlying HF
transformation. In the quantum chemistry community, a similar approach
was proposed by Koch and Dalgaard \cite{koch1993}, although the
configurations included were limited to a restricted HF-type
ansatz. The FED approach has been very successful in the nuclear
physics community (see Ref. \cite{schmid2004} and references therein).

Let us consider the variational optimization of the $n$-th determinant
in the ansatz defined by Eq. \ref{eq_mrgs:ndet}. The energy functional
becomes
\begin{align}
  {}^n E_j [ \, {}^n \Phi, \{ f \}] &= \, \frac{\langle \, {}^n
    \Psi_{j,m} | \hat{H} | \, {}^n \Psi_{j,m} \rangle}{\langle \, {}^n
    \Psi_{j,m} | \, {}^n \Psi_{j,m} \rangle}
  \nonumber \\
  &= \frac{\sum_{kl,k'l'} f^{l \ast}_k f^{l'}_{k'} \langle \, {}^l \Phi |
    \hat{H} \, \hat{P}^j_{kk'} | \, {}^{l'} \Phi
    \rangle}{\sum_{kl,k'l'} f^{l \ast}_k f^{l'}_{k'} \langle \, {}^l \Phi
    | \hat{P}^j_{kk'} | \, {}^{l'} \Phi \rangle}.
\end{align}
Here, we use the notation ${}^n E_j$ for the energy of the state to
denote that it corresponds to an $n$-determinant expansion and to
emphasize that it only depends on the label $j$ of the irreducible
representation but not on the row $m$ projected.

The variation with respect to the coefficients $f$ (note that the full
set is re-optimized) leads to the generalized eigenvalue problem
\begin{equation}
  \sum_{k'l'} \left( \, ^n \mathcal{H}_{kl,k'l'} - \, ^n E \, ^n
  \mathcal{N}_{kl,k'l'} \right) \, f^{l'}_{k'} = 0 \qquad \forall
  \qquad k, l,
\end{equation}
subject to the constraint
\begin{equation}
  \sum_{kl,k'l'} f^{l \ast}_k \, {}^n \mathcal{N}_{kl,k'l'} \,
  f^{l'}_{k'} = \delta_{kl,k'l'}.
\end{equation}
Here, the matrices $^n \mathcal{H}$ and $^n \mathcal{N}$ are given by
\begin{align}
  ^n \mathcal{H}_{kl,k'l'} &= \,
  \langle \, {}^l \Phi | \hat{H} \, \hat{P}^j_{kk'} | \, {}^{l'} \Phi
    \rangle, \\
  ^n \mathcal{N}_{kl,k'l'} &= \,
  \langle \, {}^l \Phi | \hat{P}^j_{kk'} | \, {}^{l'} \Phi \rangle.
\end{align}

A stationary point in the optimization with respect to $| \, {}^n \Phi
\rangle$ is found when
\begin{equation}
  \frac{\sum_{k,k'l'} f_k^{n \ast} \, f^{l'}_{k'} \, \langle \, {}^n
    \Phi_i^a | \left( \hat{H} - \, ^n E \right) \, \hat{P}^j_{kk'} |
    \, {}^{l'} \Phi \rangle}{\sum_{kl,k'l'} f_k^{l \ast} \,
    f^{l'}_{k'} \, \langle \, {}^l \Phi | \hat{P}^j_{kk'} | \, {}^{l'}
    \Phi \rangle} = 0 \qquad \forall \qquad i, a
\end{equation}

Because the states are constructed using a chain of variational
calculations, one can easily prove that
\begin{equation}
  ^1 E - \, ^2 E \geq \, ^2 E - ^3 E \geq \cdots \geq \, ^{n-1} E - \,
  ^n E.
\end{equation}
That is, the last added symmetry-projected configuration will improve
the ground state energy by a smaller amount than the previously added
one. Of course this is only satisfied if one can guarantee that the
global minimum was found in each optimization problem. In practice, as
this is difficult to guarantee, small deviations to this rule are
observed, yet the overall trend remains valid.

We close this section by noting that Schmid {\em et
  al}. \cite{schmid1989} realized that the FED approach is not the
most general description using $n$ symmetry-projected
configurations. The authors stated, regarding the Res HF approach
discussed in the next section, that they did not believe that ``such a
fine-tuning will yield improvements with respect to the (FED)
approach''.

\subsubsection{The resonating HF (Res HF) approach}

It is perhaps conceptually simpler, though computationally more
challenging, to optimize all configurations at the same time. This is
the basis of the resonating Hartree--Fock method devised by Fukutome
\cite{fukutome1988}. It has been used by Tomita, Ten-no, and Tanimura
\cite{tomita1996} in half-projected Res HF calculations on CO (carbon
monoxide), and by Ten-no in CI and coupled-cluster approaches based on
a Res HF expansion \cite{tenno1997}. It has proven very successful in
the context of the Hubbard model
\cite{tomita2009,tomita2004,ikawa1993,yamamoto1991}, which can be
regarded as a one-orbital-per-site cluster Hamiltonian.

Let us consider the variational optimization of the ansatz defined by
Eq. \ref{eq_mrgs:ndet}. The energy functional becomes
\begin{align}
  ^n E_j [\{ \Phi \}, \{ f \}] &= \, \frac{\langle \, {}^n \Psi_{j,m}
    | \hat{H} | \, {}^n \Psi_{j,m} \rangle}{\langle \, {}^n \Psi_{j,m}
    | \, {}^n \Psi_{j,m} \rangle}
  \nonumber \\
  &= \frac{\sum_{kl,k'l'} f^{l \ast}_k f^{l'}_{k'} \langle \, {}^l \Phi |
    \hat{H} \, \hat{P}^j_{kk'} | \, {}^{l'} \Phi
    \rangle}{\sum_{kl,k'l'} f^{l \ast}_k f^{l'}_{k'} \langle \, {}^l \Phi
    | \hat{P}^j_{kk'} | \, {}^{l'} \Phi \rangle}.
\end{align}
where we have emphasized that the full set of determinants is
optimized. Note that the form of the energy expression is the same as
in the FED approach; the difference lies in the variational
flexibility. The variation with respect to the coefficients $f$ leads
to the same generalized eigenvalue problem as in the FED approach
(though the matrix elements are necessarily different).

A stationary point in the optimization with respect to $\{ \Phi \}$ is
achieved when
\begin{equation}
  \frac{\sum_{kl,k'l'} f_k^{l \ast} \, f^{l'}_{k'} \, \langle \, {}^l
    \Phi_i^a | \left( \hat{H} - \, ^n E \right) \, \hat{P}^j_{kk'} |
    \, {}^{l'} \Phi \rangle}{\sum_{kl'',k'l'} f_k^{l'' \ast} \,
    f^{l'}_{k'} \, \langle \, {}^{l''} \Phi | \hat{P}^j_{kk'} | \,
    {}^{l'} \Phi \rangle} = 0 \qquad \forall \qquad l, i, a.
\end{equation}
This implies that the Res HF wavefunction is stationary with respect
to particle-hole mixings of any of the determinants in the expansion.

\section{Computational details}
\label{sec:details}

We have implemented the multi-component symmetry-projected HF approach
for molecular systems in an in-house program. One- and two-electron
integrals are extracted from the \verb+Gaussian 09+ \cite{gaussian}
suite. Our program is parallelized (a hybrid openMP/MPI approach is
used) over the grid-points used in the symmetry-projection as well as
over the configurations used in the multi-component expansion. Our
program is currently limited to the use of Cartesian gaussian basis
sets. We note that our FED-type implementation re-uses overlap and
Hamiltonian matrix elements and thus scales as $\mathcal{O} (n)$ with
the number of symmetry-projected configurations
\cite{rodriguez-guzman2013}.

The optimization of the broken-symmetry determinants is carried out
using a Thouless-based strategy, as described in detail in
Refs. \onlinecite{jimenez-hoyos2013,jimenez-hoyos2012b,egido1995},
with a limited-memory Quasi-Newton approach
\cite{liu1989,nocedal1980}.

One of the most important issues regarding a practical implementation
of the FED and Res HF approaches is to prepare an initial guess of the
underlying HF transformations in the symmetry-projected configuration
expansion. This was discussed in some detail by Koch and Dalgaard
\cite{koch1993}. Our approach is currently simplistic: we prepare an
initial guess of the HF transformations in the FED approach as random
unitary rotations of the orbitals closest to the Fermi energy in the
standard HF determinant or the optimized determinant in a
one-configuration symmetry-projected expansion. The unitary matrix is
built in the form $\exp (i\lambda K)$, with $\lambda \approx 0.01$ and
$K$ being a Hermitian matrix. Our initial guess for Res HF
calculations is the converged FED expansion with the same number of
determinants. Given that the symmetry-projected FED or Res HF
equations will reach a stationary point depending on the initial guess
provided, a smarter scheme to prepare the initial guess is
desirable. Nevertheless, it is difficult to anticipate {\em a priori}
the structure of general non-orthogonal determinants that will
interact the most through the Hamiltonian with the set of previously
obtained determinants.

Before we discuss our results, let us briefly clarify the nomenclature
we use. All symmetry projected methods are written in the form
X-Y. Here, Y = RHF (restricted), UHF (unrestricted), or GHF
(generalized) denotes the type of underlying HF transformation used;
complex orbitals are used in all cases. In X, we write the collection
of symmetries restored in the calculation: S is used for spin and the
point group label (like $C_{2v}$) is used to denote the type of
spatial symmetry projection.

\section{Results and discussion}
\label{sec:discussion}

\subsection{Ground-state energy of N$_2$}

\begin{figure}[!]
  \centering
  \includegraphics[width=0.85\textwidth]%
    {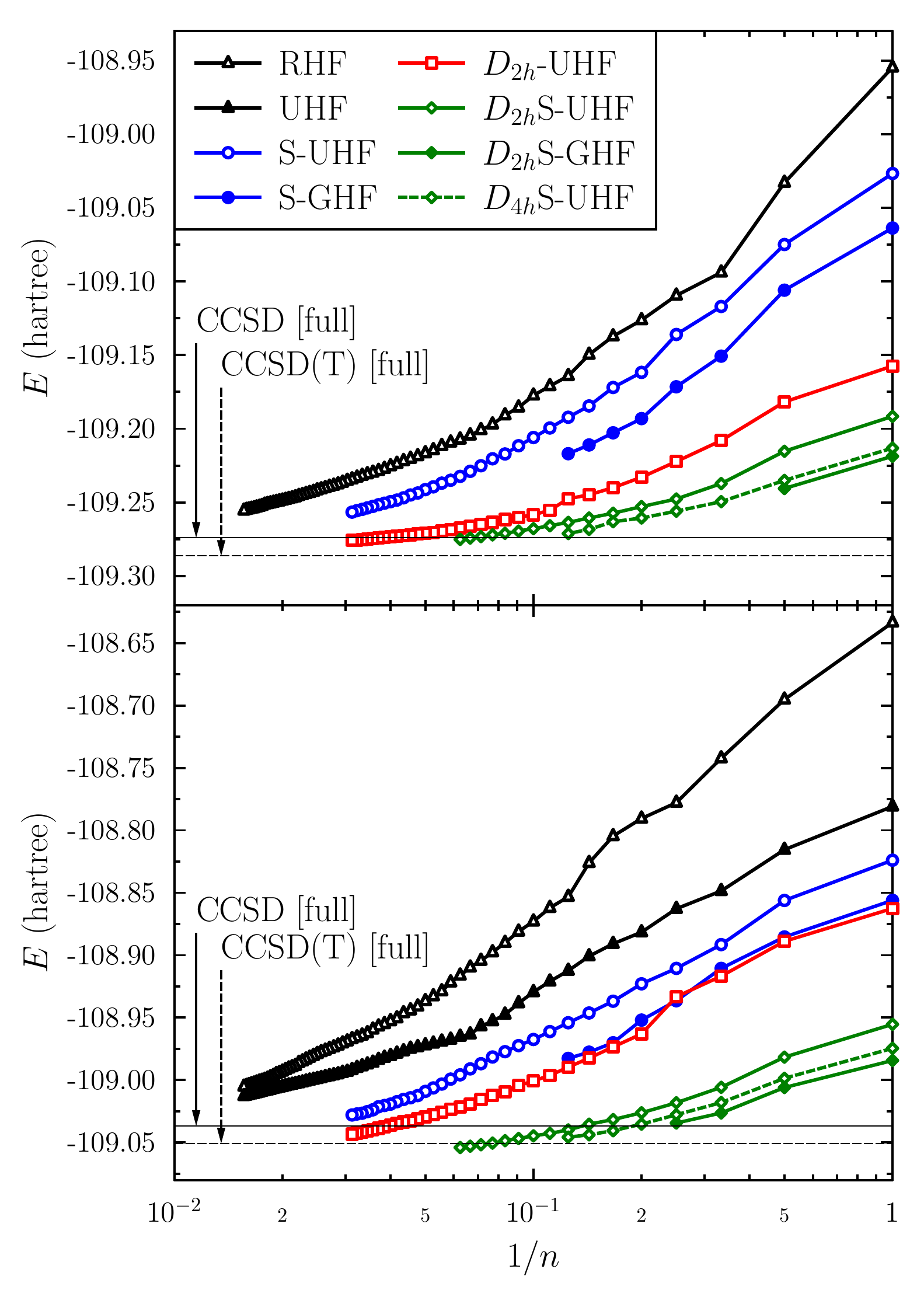}
  \caption{Ground-state energy of the N$_2$ molecule predicted by a
    variety of FED approaches at $r = r_{\mathrm{eq}}$ (top panel) and
    $r = 1.5 \, r_{\mathrm{eq}}$ (bottom panel) as a function of the
    number $n$ of symmetry-projected configurations. Here,
    $r_{\mathrm{eq}}= 1.09768 \, \mathring{\mathrm{A}}$
    \cite{nistwebbook}; a Cartesian cc-pVDZ basis set is used. The
    straight lines in each panel mark the CCSD [full] and CCSD(T)
    [full] reference energies.}
  \label{fig:n2-fed}
\end{figure}

We start by considering the ground state energy of the nitrogen
molecule, both at the equilibrium geometry ($r_{eq}$) and at the
recoupling region ($1.5 \, r_{\mathrm{eq}}$), where $r_{\mathrm{eq}} =
1.09768 \, \mathring{\mathrm{A}}$ \cite{nistwebbook}. We show in
Fig. \ref{fig:n2-fed} the evolution of the energy with the number of
transformations ($n$) added in a FED-type expansion for a series of
(symmetry-projected) methods. Calculations were performed using the
Cartesian cc-pVDZ basis set. We compare our results with
coupled-cluster singles and doubles (CCSD) [full] and CCSD(T) [full]
reference energies (obtained with the \verb+Gaussian 09+ suite). For
spin projected methods, projection to the singlet state was carried
out; for methods involving spatial symmetry projection, projection was
done to the totally symmetric irreducible representation.

We observe from the results in Fig. \ref{fig:n2-fed} that the rule
stating that the last added determinant (in FED-type expansions)
should bring less correlation than the previously added one is
satisfied in most cases. In those cases where it is not, this is
because we have failed to converge to the global minimum in the
parameter hypersurface. Several other features deserve further
discussion:
\begin{itemize}
  \item At equilibrium, spatial symmetry projection (with the $D_{2h}$
    group) brings significantly more correlation than spin projection
    with the same number of symmetry-projected configurations. This is
    not too obvious at $1.5 \, r_{\mathrm{eq}}$, yet $D_{2h}$-UHF
    remains competitive with S-GHF while being significantly cheaper.

  \item The use of broken spin-symmetry determinants (UHF-type) brings
    significantly more correlation than the use of RHF determinants at
    $1.5 \, r_{\mathrm{eq}}$. This remains true even when several
    configurations have been added; it takes roughly 4 RHF
    configurations to obtain the same energy as a single UHF
    configuration.

  \item When both spin and spatial symmetry are restored, a small
    number of configurations seems to be sufficient to obtain energies
    of comparable quality to CCSD or CCSD(T). At $1.5 \,
    r_{\mathrm{eq}}$, $\approx 16$ $D_{2h}$S-UHF configurations bring
    more correlation than CCSD(T). This is remarkable considering the
    ease of interpretation associated with the multi-component
    wavefunction.
\end{itemize}

Unfortunately, we were unable to produce such a detailed plot using
the Res HF approach, as it becomes significantly more difficult to
converge than the corresponding FED expansion. We show, nonetheless,
in Table \ref{tab:n2fedres} a comparison of ground-state energies,
evaluated at $r_{\mathrm{eq}}$, predicted with FED S-UHF and Res HF
S-UHF as a function of the number of transformations $n$.

\begin{table}[!]
  \caption{Ground-state energy of the nitrogen molecule (at $r =
    r_{\mathrm{eq}}$) predicted with the multi-component S-UHF
    approaches as a function of the number of transformations $n$. The
    Cartesian cc-pVDZ basis set was used.}
  \label{tab:n2fedres}
  \centering
  \begin{tabular}{l @{\hspace{0.5cm}} r @{\hspace{0.5cm}} r}
    \\[-18pt]
    \hline \hline \\[-18pt]
     $n$ & FED S-UHF & Res HF S-UHF \\ \hline
     1 & -109.0267 & -109.0267 \\
     2 & -109.0749 & -109.1210 \\
     3 & -109.1170 & -109.1530 \\
     4 & -109.1360 & -109.1728 \\
     5 & -109.1617 \\
     6 & -109.1720 \\
     7 & -109.1845 \\
     8 & -109.1922 \\[4pt]
     \hline \hline
  \end{tabular}
\end{table}

It is evident from the results in Table \ref{tab:n2fedres} that the
Res HF approach yields significantly lower energies than the FED
approach for a fixed number of configurations. However, the FED
approach allows one to include many more configurations than in the
Res HF approach as the optimization is cheaper and typically takes
much fewer iterations with our gradient-based optimization. For
instance, Fig. \ref{fig:n2-fed} includes results with up to 32 FED
SUHF configurations. This makes the FED approach much more convenient
for practical applications.

\subsection{Dissociation profiles}

\begin{figure}[!]
  \centering
  \includegraphics[width=0.85\textwidth]%
    {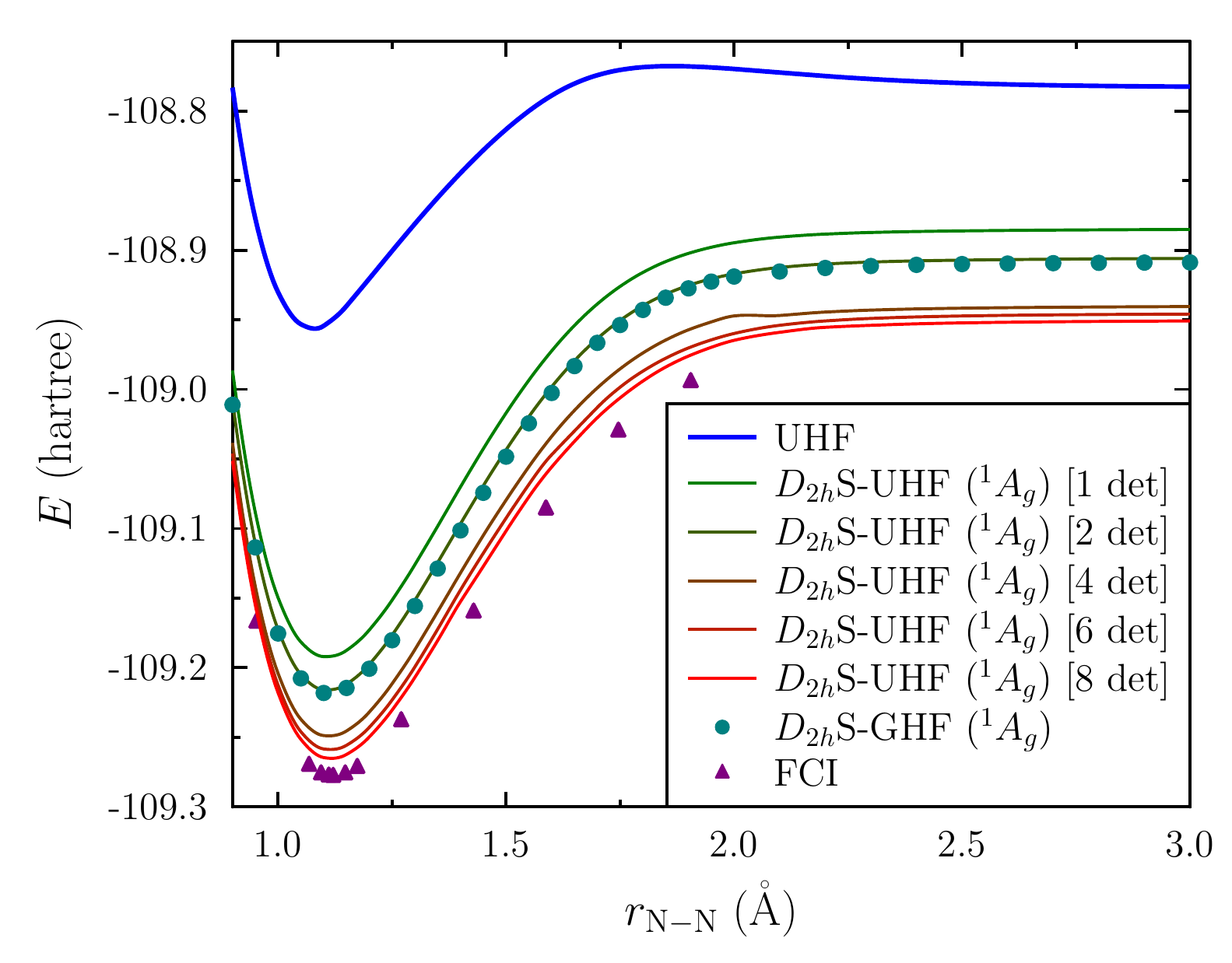}
  \caption{Dissociation profile for the ground state of the N$_2$
    molecule obtained with a FED $D_{2h}$S-UHF approach as a function
    of the number of HF transformations. The single-configuration
    $D_{2h}$S-GHF profile is included for comparison, as well as the
    FCI results from Ref. \cite{larsen2000}. The FCI results use
    spherical basis functions and frozen $1s$ orbitals (see text).}
  \label{fig:n2-fed-diss}
\end{figure}

Let us now consider the full dissociation profile of the N$_{2}$
molecule. Dissociation curves predicted with a FED $D_{2h}$S-UHF
approach are shown in Fig. \ref{fig:n2-fed-diss}, along with the
dissociation profile computed with a single symmetry-projected
configuration using $D_{2h}$S-GHF. The calculations use the Cartesian
cc-pVDZ basis set. We compare our curves with the FCI profile from
Ref. \cite{larsen2000}. Nevertheless, we stress that the FCI results
are not directly comparable: they were obtained with the spherical
cc-pVDZ basis and freezing the $1s$ core orbital of the N atoms. Both
of these effects would contribute to underestimate the FCI
energy \footnote{Near $r_{\mathrm{eq}}$, CCSD yields a 12.3 mhartree
  energy difference once both effects are accounted for.}. The FCI
results are included as a guide to the eye for the correct shape of
the dissociation curve.

The results in Fig. \ref{fig:n2-fed-diss} show that $D_{2h}$S-UHF
yields a qualitatively correct dissociation curve even with a single
symmetry-projected configuration. Inclusion of 8 symmetry-projected
configurations (using the FED approach) results in a curve fairly
parallel to the reference FCI curve. The energy improvement due to the
additional configurations is seen accross the potential energy
surface. Interestingly, 2 symmetry-projected configurations with the
$D_{2h}$S-UHF method match the results from $D_{2h}$S-GHF with a
single configuration. Finally, let us stress that the wavefunction
resulting from a multi-component approach can be regarded as a
discretized form of the unitary group coherent state representation of
the exact many-fermion state \cite{linderberg1977}. Therefore, in the
limit of a large number of configurations, the size-consistency error
associated with projected HF approaches necessarily disappears.

\begin{figure}[!h]
  \centering
  \includegraphics[width=0.85\textwidth]%
    {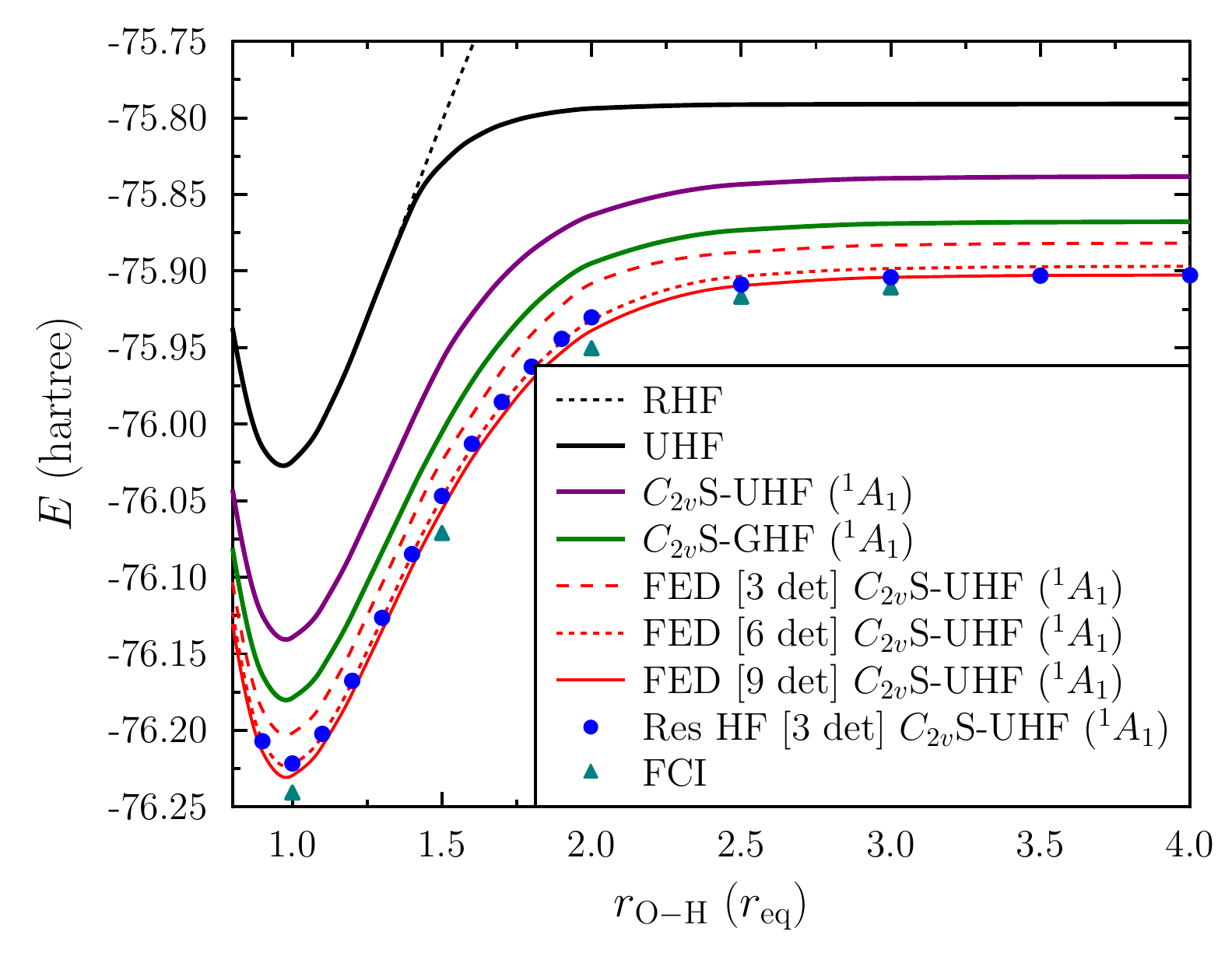}
  \caption{Symmetric dissociation profile of the water molecule as
    predicted with a variety of multi-configuration symmetry-projected
    approaches. A Cartesian cc-pVDZ basis set was used. FCI results
    from Ref. \cite{olsen1996}. The FCI results use spherical basis
    functions (see text).}
  \label{fig:h2o-fed-diss}
\end{figure}

We consider in Fig. \ref{fig:h2o-fed-diss} the symmetric dissociation
profile of the H$_2$O molecule, as predicted with a variety of
multi-configuration symmetry-projected approaches. The Cartesian
cc-pVDZ basis set was used in our calculations, whereas the FCI
results from Ref. \cite{olsen1996} were obtained with the spherical
cc-pVDZ basis set \footnote{At $r_{\mathrm{eq}}$, CCSD is 3.5 mhartree
  lower in the Cartesian basis set.}. The restored quantum numbers in
symmetry-projected calculations are $s=0$ for spin and the $A_1$
irreducible representation of the $C_{2v}$ group.

Results from Fig. \ref{fig:h2o-fed-diss} show a similar scenario as
that observed in the N$_2$ dissociation. The $C_{2v}$S-UHF
dissociation profile is already fairly parallel to the FCI
solution. It accounts for significantly more correlations near
equilibrium than towards dissociation with respect to the UHF
solution. With 9 symmetry-projected configurations, the FED
$C_{2v}$S-UHF curve is only a few mhartree off from the FCI curve
across the entire potential energy surface.

We have been able to compute the entire dissociation profile using a
3-configuration Res HF approach based on $C_{2v}$S-UHF. Interestingly,
it yields similar results as a 6-configuration FED approach near
equilibrium, but becomes more accurate towards dissociation, rivaling
the 9-configuration FED approach. This is a result of the increased
flexibility in the Res HF ansatz. Finally, Fig. \ref{fig:h2o-fed-diss}
also shows the dissociation profile predicted with the $C_{2v}$S-GHF
method, using a single configuration. Quite disappointingly, the
results are only comparable to a two-configuration $C_{2v}$S-UHF
wavefunction, even though the former is almost two orders of magnitude
more expensive to evaluate.

\subsection{The copper oxide [Cu$_2$O$_2$]$^{2+}$ core}

We have recently applied the projected Hartree--Fock method
\cite{samanta2012} to the theoretical study of the copper oxide cores,
in particular, the interconversion profile between the
$\mu-\eta^2:\eta^2$-peroxodicopper(II) ($\mathbf{A}$) and the
bis($\mu$-oxo)-dicopper(II) ($\mathbf{B}$) forms.
\begin{figure}[!h]
  \centering
  \includegraphics[width=0.5\textwidth]{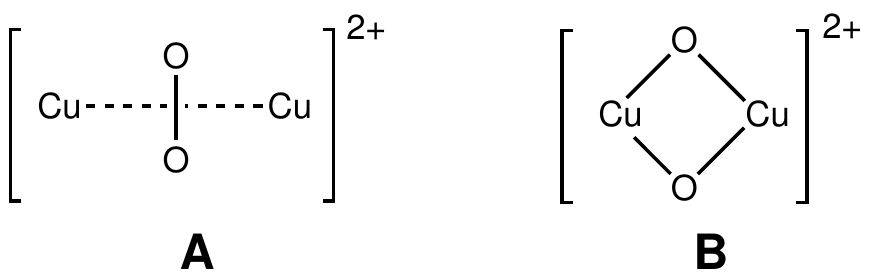}
  \caption{Structures of $\mu-\eta^2:\eta^2$-peroxodicopper(II)
    ($\mathbf{A}$) and bis($\mu$-oxo)-dicopper(II) ($\mathbf{B}$) in
    the interconversion profile of the [Cu$_2$O$_2$]$^{2+}$ core.}
\end{figure}

The interconversion profile of the bare [Cu$_2$O$_2$]$^{2+}$ core has
been recently studied theoretically by Cramer {\em et
  al}. \cite{cramer2006}, Malmqvist {\em et al}. \cite{malmqvist2008},
and Yanai {\em et al}. \cite{yanai2010} with a variety of highly
sophisticated ab initio methods. This system has proven tremendously
challenging due to the expected multi-reference character in
$\mathbf{A}$ and the large active space that one has to include in
traditional multi-reference approaches (a reasonable active space for
this system would involve 30 electrons in 28 orbitals).

It should be pointed out that recently Liakos and Neese
\cite{liakos2011} have shown that the multi-reference character in the
copper oxide core is very limited. They examined the influence of
ligands as well as relativistic and solvent effects and concluded that
the single-reference based local-pair natural orbital coupled-cluster
method in fact provides very reliable profiles for this system. Their
assessment is likely valid in the presence of ligands and solvent, and
is hence relevant for comparison with experimental results. On the
other hand, we can still treat the bare copper oxide core as a toy
system for which different highly sophisticated theoretical methods
yield inconsistent results.

\begin{figure}[!htbp]
  \centering
  \includegraphics[width=0.85\textwidth]%
    {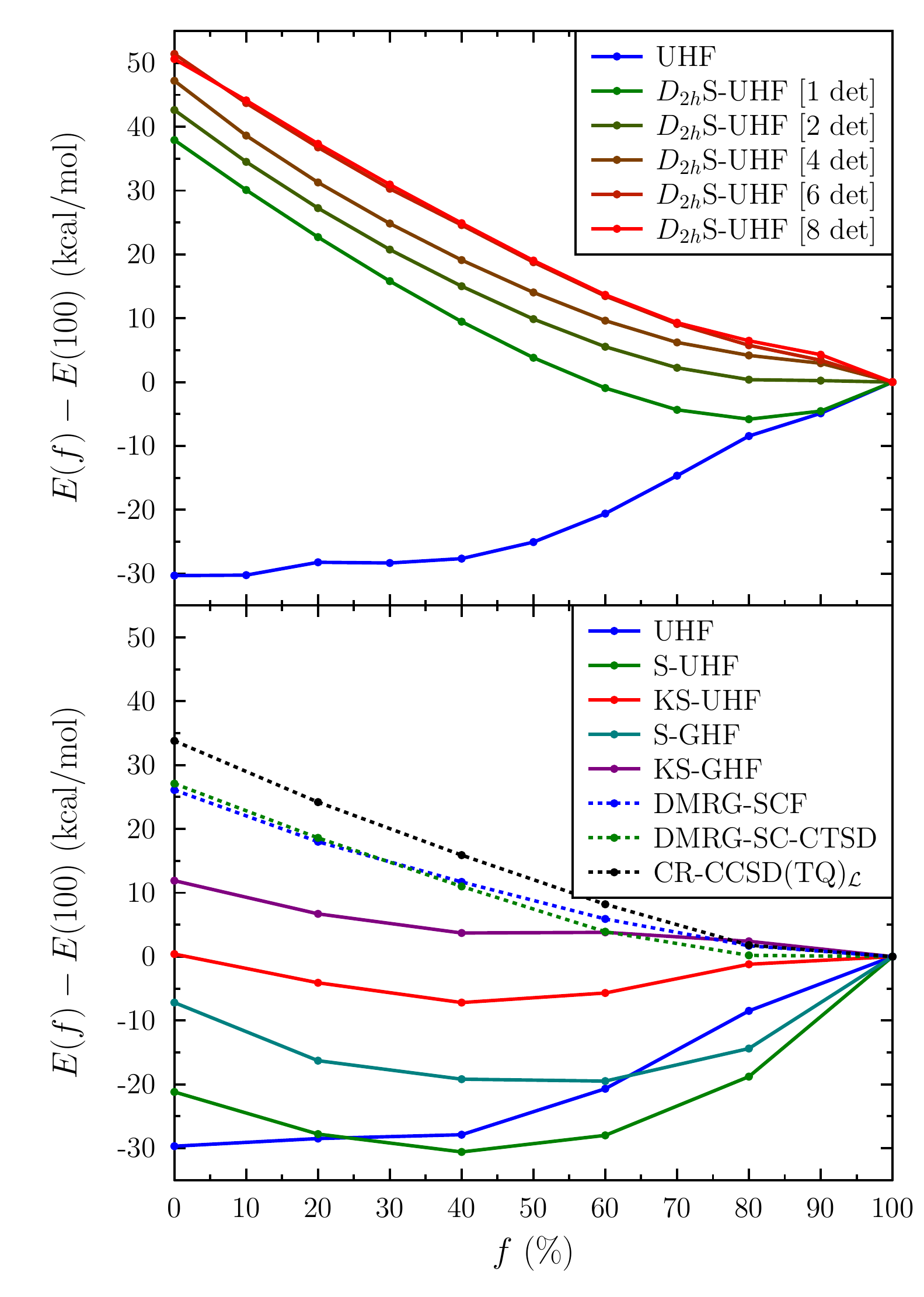}
  \caption{Relative total energy of [Cu$_2$O$_2$]$^{2+}$ along a
    linear isomerization path from $\mathbf{B}$ ($f=0$) to
    $\mathbf{A}$ ($f=100$). A variety of single symmetry-projected
    configuration methods are shown in the bottom panel, while a
    multi-reference FED approach based on $D_{2h}$S-UHF is shown in
    the top panel. CR-CC data was extracted from
    Ref. \cite{cramer2006}, while DMRG data was taken from
    Ref. \cite{yanai2010}. The upper panel uses a Cartesian basis set
    while the lower panel uses spherical basis functions (see text).}
  \label{fig:cu2o2}
\end{figure}

In Ref. \cite{samanta2012} we assessed the ability of single-reference
symmetry-projected methods to accurately describe the interconversion
profile of the bare copper oxide core. Understanding that the RASPT2
(restricted active-space second order perturbation theory)
\cite{malmqvist2008}, CR-CC (completely renormalized coupled-cluster)
\cite{cramer2006}, and DMRG-SC-CTSD (density-matrix renormalization
group with strongly contracted canonical transformation including only
single and double excitations) \cite{yanai2010} methods provide the
likely correct profile for this system, we ranked the S-UHF, S-GHF,
KS-UHF, and KS-GHF methods according to how close they came to the
former methods. We observed that the more symmetries restored the
closer the profile got to the reference methods. We show, in the lower
panel of Fig. \ref{fig:cu2o2}, a summary of the results presented in
Ref. \cite{samanta2012}.

We have revisited our results for the interconversion profile of the
copper oxide core with our multi-reference FED approaches. We have
employed the same basis set as our previous work, save for the fact
that our program cannot currently handle spherical basis sets. The
effect of the change of basis is expected to be very small and should
not affect the conclusions of our work \footnote{The use of Cartesian
  basis functions lowers the UHF energy at $f=100$ by 3.2 mhartree.}.
The totally symmetric irrep of the $D_{2h}$ group was restored in our
calculations.

The upper panel of Fig. \ref{fig:cu2o2} shows the interconversion
profiles obtained by our FED $D_{2h}$S-UHF approach as a function of
the number of transformations included. We note that the restoration
of spatial symmetry makes a huge difference even when a single
configuration is included. A single-determinant $D_{2h}$S-UHF approach
predicts $\mathbf{A}$ to be $\approx 38$ kcal/mol higher in energy
than $\mathbf{B}$, and the profile closely resembles those that we
have deemed as accurate. Increasing the number of transformations
further raises the energy of $\mathbf{A}$ relative to
$\mathbf{B}$. Our interconversion profile seems to converge with 6
symmetry-projected configurations to a relative energy of $\approx 50$
kcal/mol, notably higher than the RASPT2, CR-CC, and DMRG-SC-CTSD
curves. Because our results show a relatively smooth convergence with
the number of configurations, we believe our results could be more
accurate than the ones just quoted.

\section{Conclusions}
\label{sec:conclusions}

In this work we have considered a multi-component approach to account
for the correlations missing in the symmetry-projected ansatz for the
ground state of a molecular system. The ground state description is
improved by making a linear combination of symmetry-projected
configurations constructed from a set of (generally non-orthogonal)
deformed Slater determinants. Two extreme optimization strategies were
considered: a FED approach where only the last-added determinant is
optimized (along with the full set of linear variational
coefficients), and a Res HF approach where all the variational
parameters are optimized at once.

We note that our multi-component approach is exact in the limit of an
infinite number of symmetry-projected configurations included in the
expansion, regardless of the optimization strategy used. In such
limit, the wavefunction coincides with the coherent state
representation of the exact wavefunction
\begin{equation}
  |\Psi \rangle = \Bigg( \int \prod_{ph} dz_{ph} \, dz^\ast_{ph} \,
    \mu(\mathbf{z}) \, |\mathbf{z} \rangle \, \langle \mathbf{z}
    |\Bigg) | \Psi \rangle,
\end{equation}
where $|\mathbf{z} \rangle$ is a generalized fermion coherent state
\cite{gazeau} generated from a Thouless rotation out of a reference
Slater determinant $|\Phi_0 \rangle$ (see, {\em e.g.},
Ref. \onlinecite{blaizot_ripka}). Here, $\mu (\mathbf{z})$ is a
measure guaranteeing that the closure relation (the term in
parenthesis) equals the identity operator.

Our work has shown that for molecular systems a FED approach tends to
be more efficient than a Res HF one in building ground state
correlations, even if the latter yields a more elegant
wavefunction. This is because convergence is easier and the
optimization problem can be implemented with linear computational cost
in the number of transformations. We have observed that a few
symmetry-projected configurations are sufficient to account for most
of the correlations (both weak and strong) in simple molecular
systems, such as the nitrogen and the water molecule. Near
equilibrium, we can even obtain variational energies (in small basis
sets) that are near the coupled-cluster ones.

In addition, we have revisited the copper oxide cores as an example of
a challenging multi-reference system where both static and dynamic
correlations are significant. By using a FED expansion in terms of
symmetry-projected configurations with good spatial and spin
symmetries we were able to improve our results yielding a linear
isomerization path that is of comparable quality as those previously
reported with RASPT2, CR-CC, or DMRG-SC-CTSD.

An interesting question that results from this work is to determine
the most efficient prescription to account for these correlations. We
have observed, for instance, that two S-UHF configurations tend to
give energies that are of similar quality as a single S-GHF
configuration, while the latter involves a computational effort that
is roughly two orders of magnitude larger because of the size of the
respective integration grids. This need not, however, be true for all
systems: a frustrated configuration such as an equilateral H$_3$
triangle will undoubtedly benefit significantly from the use of
non-collinear deformed determinants. Given a determinantal expansion
of a fixed length, letting all the determinants be independent will
always afford the best description. Nevertheless, by constructing a
same-size expansion in terms of the superposition of the Goldstone
manifolds of fewer broken-symmetry states one may obtain a
wavefunction that is near in quality to the former one. The latter has
the virtue of respecting all symmetries of the Hamiltonian and being
defined by a smaller number of computational parameters, thus becoming
easier to optimize. Identifying those ``efficient symmetries'' is
certainly of paramount importance for practical applications.

\section*{Acknowledgments}
\label{sec:acknow}

This work was supported by the Department of Energy, Office of Basic
Energy Sciences, Grant No. DE-FG02-09ER16053. G.E.S. is a Welch
Foundation Chair (C-0036). C.A.J.H. acknowledges support from the
Lodieska Stockbridge Vaughn Fellowship.

\appendix

\section{Matrix elements between symmetry-projected configurations}
\label{sec:matrix_elements}

In this appendix we provide explicit expressions for matrix elements
between symmetry-projected configurations. We note that these can be
expressed in terms of matrix elements between non-orthogonal Slater
determinants, for which an extended Wick's theorem can be used as
shown by, {\em e.g.}, Blaizot and Ripka \cite{blaizot_ripka}. A
detailed derivation of the form of the matrix elements can be found
in, {\em e.g.}, Ref. \onlinecite{jimenezthesis}.

We assume a non-relativistic, Born-Oppenheimer molecular electronic
Hamiltonian $\hat{H}$ expressed in the form
\begin{equation}
  \hat{H} = \sum_{ik} \langle i | \hat{h} | k \rangle \, c_i^\dagger
    \, c_k + \frac{1}{4} \sum_{ijkl} \langle ij | \hat{v} | kl \rangle
    \, c_i^\dagger \, c_j^\dagger \, c_l \, c_k,
\end{equation}
where $\langle i | \hat{h} | k \rangle$ are one-electron (core
Hamiltonian) integrals and $\langle ij | \hat{v} | kl \rangle$ are
anti-symmetrized two-electron (electron repulsion) integrals in Dirac
notation. In addition, we use the matrix $D^\ast$ to relate the
orbitals in a Slater determinat to the basis states (assumed, without
loss of generality, to be orthonormal).

Overlap and Hamiltonian matrix elements between symmetry-projected
configurations are expressed in terms of norm and Hamiltonian overlaps
between rotated determinants $\hat{R} (\vartheta) |\Phi \rangle$ as
\begin{subequations}
  \begin{align}
    \langle \, {}^r \Phi | \hat{P}^j_{kk'} | \, {}^s \Phi \rangle
    &= \, \frac{1}{V} \int_V d\vartheta \, w^j_{kk'} (\vartheta) \,
      n^{rs} (\vartheta), \\[4pt]
    \langle \, {}^r \Phi | \hat{H} \, \hat{P}^j_{kk'} | \, {}^s \Phi
      \rangle
    &= \, \frac{1}{V} \int_V d\vartheta \, w^j_{kk'} (\vartheta) \,
      n^{rs} (\vartheta) \, h^{rs} (\vartheta),
  \end{align}
\end{subequations}
where
\begin{subequations}
  \begin{align}
    n^{rs} (\vartheta) &\equiv \,
    \langle \, {}^r \Phi | \hat{R} (\vartheta) | \, {}^s \Phi \rangle,
    \label{eq:def_nrs} \\[4pt]
    h^{rs} (\vartheta) &\equiv \,
    \frac{\langle \, {}^r \Phi | \hat{H} \, \hat{R} (\vartheta) | \,
      {}^s \Phi \rangle}{\langle \, {}^r \Phi | \hat{R} (\vartheta) |
      \, {}^s \Phi \rangle}.
    \label{eq:def_hrs}
  \end{align}
\end{subequations}

The norm overlaps of Eq. \ref{eq:def_nrs} can be evaluated with
\begin{align}
  n^{rs} (\vartheta) &= \, \mathrm{det}_N \, X^{rs} (\vartheta), \\
  X^{rs} (\vartheta) &= \, D^{r\trans} \, R(\vartheta) \, D^{s\ast}.
  \label{eq:def_xrs}
\end{align}
Here, the notation $\mathrm{det}_N$ is used to emphasize that the
determinant should be evaluated over the $N \times N$ block of $X^{rs}
(\vartheta)$. That is, only the rectangular matrices $D^k$ of occupied
orbitals should be used in the expression above. In addition, $R
(\vartheta)$ is the matrix representation of the rotation operator
$\hat{R} (\vartheta)$ in the single-particle basis.

The Hamiltonian overlaps of Eq. \ref{eq:def_hrs} are given by
\begin{align}
  h^{rs} (\vartheta) &= \, \sum_{ik} \left[ \langle i | \hat{h} | k
    \rangle + \frac{1}{2} \, \Gamma^{rs}_{ik} (\vartheta) \right]
    \rho^{rs}_{ki} (\vartheta), \\
  \Gamma^{rs}_{ik} (\vartheta) &= \, \sum_{jl} \langle ij | \hat{v} |
    kl \rangle \, \rho^{rs}_{lj} (\vartheta),
\end{align}
where we have expressed them in terms of the transition density matrix
$\rho^{rs} (\vartheta)$. The latter can be built according to
\begin{equation}
  \rho^{rs} (\vartheta) = R(\vartheta) \, D^{s\ast} \, \left[
    X^{rs} (\vartheta) \right]^{-1} \, D^{r\trans}.
\end{equation}
Here, the inverse of $X^{rs} (\vartheta)$ (defined in
Eq. \ref{eq:def_xrs}) should be evaluated over the $N \times N$ block
of occupied orbitals in both determinants.

Matrix elements appearing in contributions to the energy gradient can
also be expressed in terms of overlaps between rotated determinants:
\begin{subequations}
  \begin{align}
    \langle \, {}^r \Phi_h^p | \hat{P}^j_{kk'} | \, {}^s \Phi \rangle
    &= \, \frac{1}{V} \int_V d\vartheta \, w^j_{kk'} (\vartheta) \,
      n^{rs} (\vartheta) \, N^{rs}_{ph} (\vartheta), \\[4pt]
    \langle \, {}^r \Phi_h^p | \hat{H} \, \hat{P}^j_{kk'} | \, {}^s
      \Phi \rangle
    &= \, \frac{1}{V} \int_V d\vartheta \, w^j_{kk'} (\vartheta) \,
      n^{rs} (\vartheta) \, H^{rs}_{ph} (\vartheta).
  \end{align}
\end{subequations}
Here,
\begin{subequations}
  \label{eq:grad_rs}
  \begin{align}
    N^{rs}_{ph} (\vartheta) &\equiv \,
    \frac{\langle \, {}^r \Phi_h^p | \hat{R} (\vartheta) | \, {}^s
      \Phi \rangle}{\langle \, {}^r \Phi | \hat{R} (\vartheta) | \,
      {}^s \Phi \rangle},
    \\[4pt]
    H^{rs}_{ph} (\vartheta) &\equiv \,
    \frac{\langle \, {}^r \Phi_h^p | \hat{H} \, \hat{R} (\vartheta) |
      \, {}^s \Phi \rangle}{\langle \, {}^r \Phi | \hat{R} (\vartheta)
      | \, {}^s \Phi \rangle}.
  \end{align}
\end{subequations}

The matrix elements of Eq. \ref{eq:grad_rs} are given by
\begin{subequations}
  \begin{align}
    N^{rs}_{ph} (\vartheta) &= \,
    \left[ D^{r\trans} \, \rho^{rs} (\vartheta) \, D^{r\ast}
      \right]_{ph},
    \\[4pt]
    H^{rs}_{ph} (\vartheta) &= \,
    h^{rs} (\vartheta) \left[ D^{r\trans} \, \rho^{rs} (\vartheta) \,
      D^{r\ast} \right]_{ph}
    \nonumber \\
    &+ \, \left[ D^{r\trans} \, \left( \mathbf{1} -
      \rho^{rs} (\vartheta) \right) \, f^{rs} (\vartheta) \,
      \rho^{rs} (\vartheta) \, D^{r\ast} \right]_{ph},
  \end{align}
\end{subequations}
where we have set $f^{rs}_{ik} (\vartheta) = \langle i | \hat{h} | k
\rangle + \Gamma^{rs}_{ik} (\vartheta)$.

%

\end{document}